\documentclass[%
 aip,
 amsmath,amssymb,
reprint, 
]{revtex4-1}

\usepackage{graphicx}
\usepackage{dcolumn}
\usepackage{bm}

\usepackage[utf8]{inputenc}
\usepackage[T1]{fontenc}
\usepackage{mathptmx}
\usepackage{etoolbox}
\usepackage{amsmath}
\usepackage{amssymb}
\usepackage[colorlinks=true,linkcolor=blue]{hyperref}
\hypersetup{
  colorlinks = true,
  citecolor = blue,
  urlcolor = blue
}
\newcommand{\R}{\mathcal{R}}
\newcommand{\F}{\mathcal{F}}
\newcommand{\vp}{\bm{v}_\perp}
\newcommand{\bp}{\bm{b}_\perp}
\newcommand{\bb}{\bm{b}}
\newcommand{\bnabla}{\bm{\nabla}}
\newcommand{\rb}{\bm{r}}
\newcommand{\wz}{\bm{\omega}}
\newcommand{\jz}{\bm{j}}
\newcommand{\cross}{\bm{\times}}
\newcommand{\la}{\left\langle}
\newcommand{\ra}{\right\rangle}
\newcommand{\lbracket}{\left[}
\newcommand{\rbracket}{\right]}
\makeatletter
\def\@email#1#2{%
 \endgroup
 \patchcmd{\titleblock@produce}
  {\frontmatter@RRAPformat}
  {\frontmatter@RRAPformat{\produce@RRAP{*#1\href{mailto:#2}{#2}}}\frontmatter@RRAPformat}
  {}{}
}%
\makeatother
\begin{document}


\title{Universal cascade and relaxation of strong anisotropic turbulence in fusion plasmas}
\author{Ramesh Sasmal}
\author{Supratik Banerjee}%
 \email{sbanerjee@iitk.ac.in}
\affiliation{ 
Department of Physics, Indian Institute of Technology Kanpur, Uttar Pradesh 208016, India
}%


\begin{abstract}
Starting from the governing equations, exact relations have been derived for three-dimensional reduced magnetohydrodynamic turbulence corresponding to the inertial range cascade of energy and cross-helicity. Justifications are provided for not attempting to recover the said exact relations as a limit of the exact relations previously derived for incompressible magnetohydrodynamic turbulence. Assuming axial symmetry, anisotropic energy spectrum has been predicted from the exact relation and is found to be consistent with the critical balance thus leading to a $-5/3$ perpendicular energy spectrum. In the case of a strong alignment between the velocity and the magnetic field fluctuations, the derived exact relation implies a generalized anisotropic spectrum with a $-3/2$ power-law dependence in the direction of alignment. Using the alternative form of the exact relations, it is shown that the flow naturally relaxes towards a state of dynamic alignment in the limit of negligible kinetic and magnetic pressure. Finally, despite having different equations of dynamics, the exact relations for energy and cross-helicity and the relaxed states of a two-dimensional MHD are found to be identical to those in reduced magnetohydrodynamic flow. 
\end{abstract}

\maketitle

\section{Introduction}

A fluid flow becomes turbulent at very large Reynolds number associated with dominating nonlinearity over the viscous effects. The understanding of turbulence is necessary to explain the efficient mixing of different parts of a flow, structure formation, and rapid heating of the flow medium. For the length scales superior to the ion inertial length, plasmas can be described using a single magnetohydrodynamic (MHD) fluid. A turbulent regime can be obtained for an MHD fluid when both the Reynolds and the magnetic Reynolds numbers are very high. In a fully developed turbulence, energy nonlinearly cascades (stepwise) from the largest flow scales to the smallest ones where finally it gets dissipated. In particular, there exists an intermediate range of scales, called the inertial range, where the energy cascades with a constant (independent of scales) transfer rate $\varepsilon$. This $\varepsilon$, being independent of the flow geometry and the fluid viscosity, represents the universality of turbulence. Under the assumption of homogeneity and isotropy, such universality can be characterized by a $E(k)\sim k^{-5/3}$ energy spectrum for hydrodynamic (HD) turbulence that follows Kolmogorov phenomenology where the energy cascades due to the fragmentation of eddies \citep{kolmogorov1941}. For isotropic MHD turbulence, when the mean magnetic field (${\bf B_0}$) is strong,  an energy spectrum $E(k)\sim k^{-3/2}$ is obtained following Iroshnikov-Kraichnan phenomenology (IK) where the energy cascades to the smaller scales due to the sporadic interactions of counter-propagating Alfv\'en waves \citep{iroshnikov1964turbulence, kraichnan1965inertial}. However, unlike a mean velocity field, ${\bf B_0}$ cannot be eliminated by a Galilean transformation and hence, MHD turbulence becomes anisotropic in the presence of a background magnetic field. The turbulent fluctuations mainly permeate the plane perpendicular to ${\bf B_0}$ whereas the ensemble of the fluctuations are mainly advected by the mean field along it. As it is expected, the turbulent power transfer along the mean field is often dominated by that in the plane perpendicular to the mean field. When the anisotropy is such that the linear Alfv\'en time ($\tau_A\sim 1/ k_\parallel B_0$) is comparable to that of the nonlinear time ($\tau_\ell\sim 1/k_\perp b_\perp$), a regime of critical balance is achieved and corresponding anisotropic power spectrum is given by $E (k_\perp, k_\parallel) \sim k_\perp^{-5/3} k_\parallel^{-1}$ \citep{goldreich1995}. In this regime, using the constitutive relation $k_\parallel \sim k_\perp^{2/3}$, one can separately obtains the perpendicular power spectrum $E(k_\perp)\sim k_\perp^{-5/3}$ and a parallel power spectrum $E (k_\parallel)\sim k_\parallel^{-2}$. If $\tau_A \ll \tau_\ell$, one practically encounters the anisotropic IK situation and obtains $E(k_\perp, k_\parallel)\sim k_\perp^{-2} k_\parallel^{-1/2}$. If the guiding magnetic field is very strong, the turbulence can effectively be assumed to take place in the plane perpendicular to $\bf B_0$. A satisfactory theoretical description can be obtained by the equations of reduced magnetohydrodynamics (RMHD) which was initially derived heuristically to model tokamak plasmas \citep{kadomtsev1973nonlinear, strauss1976nonlinear}. Both the pioneering works are based on nearly incompressible low-$\beta$ MHD plasma. However, \citet{kadomtsev1973nonlinear} performed an order analysis using $B_\perp/B_0\sim\varepsilon$ as the smallness parameter whereas the wave vector anisotropy $ k_\parallel/k_\perp \sim\varepsilon$ was used in the analysis by \citet{strauss1976nonlinear}. This apparent discrepancy can be reconciled since RMHD is a strong turbulence model where $\tau_\ell\sim\tau_A$. In the RMHD limit, both these nearly incompressible models should be free from fast magnetosonic waves which is warranted if $v_\parallel$ and $b_\parallel$ are negligibly small. Later, the same set of equations was also derived using a more rigorous perturbative analysis of both three-dimensional incompressible and nearly incompressible MHD equations \citep{montgomery1982, zank1992equations}. Interestingly, \citet{montgomery1982} uses incompressible MHD equations which are free from magnetosonic waves even for nonnegligible parallel fluctuations and hence, the corresponding RMHD equations are obtained from the perpendicular components of the MHD equations. The RMHD equations were also recovered as a fluid limit of gyrokinetics for length scales much larger than the ion gyroradius \citep{schekochihin2009astrophysical}.

Besides the context of fusion plasmas, the RMHD model has also been studied in the realm of space and astrophysical plasmas, including solar wind turbulence, coronal heating, reconnection, and coronal loop dynamics \citep{milano1999, oughton2001astro, rappazzo2007coronal,dalena2014test}. Several numerical studies of RMHD turbulence have also been accomplished in the past few decades. For boundary-driven RMHD turbulence, a perpendicular power spectrum $E (k_\perp) \sim k_\perp^{-5/3}$ is observed for $\tau_A /\tau_\ell=2$ \citep{dmitruk2003energy}. Using a parallel simulation of RMHD, a similar perpendicular spectrum has also been obtained for equal magnetic and kinetic dissipation \citep{gomez2005parallel}. In a comparative study between compressible 3D MHD and RMHD, $E (k_\perp) \sim k_\perp^{-5/3}$ has also been found in the inertial range \citep{dmitruk2005direct}. Despite a large number of numerical studies on the perpendicular energy spectra,  no systematic analytical study has been carried out on the inertial range cascades of RMHD turbulence. In addition, it is also interesting to investigate the relaxed states when these cascades are quenched. 

In this paper, we derive two exact relations, one with classical divergence form \citep{podesta2008, banerjee2013exact} and the other with the alternative form \citep{banerjee2016alternative, banerjee2016chiral, banerjee2018energy}, for the inertial range energy cascade in the three-dimensional RMHD turbulence. From the divergence form, we also predict a $k_\perp^{-5/3}$ power law for the perpendicular energy spectra which has been observed in several numerical studies mentioned above. Using the alternative form along with the recently proposed universal theory of fluids and plasma relaxation \citep{banerjee2023universal, pan2024universal}, we show that a turbulent RMHD system relaxes to a state of Alfv\'enic alignment where $\bp=\pm\vp$ with $\bp$ and $\vp$ being the magnetic field fluctuation (in the Alfv\'en unit) and velocity fluctuation respectively in the plane perpendicular to $\bm{b_0}$. Finally, we compare RMHD with two-dimensional MHD in terms of turbulent energy transfer.  

This paper is organized as follows: in Sec. \ref{model}, we give the governing equations of the RMHD model followed by the total energy and cross helicity conservation in the inviscid limit. Sec. \ref{exact relation} consists of the detailed derivation of the exact relation in divergence form along with a power law prediction for the perpendicular energy spectra. In Sec. \ref{alternative exact relation}, the alternative form of exact relation is derived and hence, the turbulent relaxed state of RMHD is obtained using recently proposed PVNLT. This is followed by Sec. \ref{comparision} where we compare RMHD turbulence with two-dimensional MHD turbulence. Finally, in Sec. \ref{discussion}, we summarize our results and conclude.

\section{RMHD model}\label{model}
\subsection{Governing equations}
As discussed above, the ordinary MHD equations reduce to the RMHD equations in the presence of a uniform mean magnetic field $\bm{b_0}$ which is considerably stronger than the velocity and magnetic field fluctuations. Under such situation, the flow becomes anisotropic and the total gradient operator $\bnabla$  should be written as a sum of $\bnabla_\parallel$ (gradient along $\bm{b_0}$) and $ \bnabla_\perp $ (gradient perpendicular to $\bm{b_0}$). In particular, the fluctuation length scale parallel to $\bm{b_0}$ becomes larger than that perpendicular to $\bm{b_0}$ thus requiring $k_\parallel <k_\perp$ where $\bnabla_\parallel \sim k_\parallel \hat{b_0} $ and $\bm{k_\perp} \sim \bnabla_\perp  $. 
If the magnetic field  $\bm{b}= b_0 \hat{z} + \bp$ and the velocity field  $\bm{v}=\vp$, then
the equations of RMHD are written as \citep{zank1992equations, oughton2017reduced2dor3d}
{\small
\begin{align}
    \frac{\partial \vp}{\partial t} + (\vp\cdot\bnabla_\perp) \vp &= -\bnabla_\perp p^*+(\bp\cdot\bnabla_\perp) \bp \nonumber
    \\&+b_0\frac{\partial \bp}{\partial z}+ \bm{f}+\nu \bnabla_\perp^2\vp,\label{zmc}\\
    \frac{\partial \bp}{\partial t}+(\vp\cdot\bnabla_\perp) \bp&=\left(\bp\cdot\bnabla_\perp\right) \vp+b_0\frac{\partial \vp}{\partial z}\nonumber\\
    &+\eta\bnabla_\perp^2\bp,\label{zmi}\\
    \bnabla_\perp\cdot\vp&=0,\;\; \bnabla_\perp\cdot\bp=0,\label{zm2}
\end{align}}

\noindent where  $\nu$, $\eta$, $\bm{f}$, and $p^*$ denote the fluid viscosity,  magnetic resistivity, turbulent forcing, and total pressure (fluid pressure ($p$) plus magnetic pressure ($\bp^2/2$)) respectively. The RMHD equations, originally derived by \citet{strauss1976nonlinear}, were in terms of stream function (U) and the magnitude of the vector potential (A). However, the above equations can be recovered from them by writing  $\vp=\bnabla U\cross\hat{\bm{z}
}$ and $\bp=\bnabla A\cross\hat{\bm{z}}$.

\subsection{Inviscid invariants}
In this part, we search for the quadratic inviscid invariants of an RMHD flow.
For ideal MHD, the total energy, cross helicity, and magnetic helicity are three conserved quantities \citep{banerjee2014compressible}. By construction (as mentioned above), the vector potential in RMHD is given by $\bm{A}= A \hat{z}$ and hence the magnetic helicity density ($\bp \cdot \bm{A}$) identically vanishes at every point. 

Similar to MHD, in RMHD, one can define the total energy ($E$) and the cross helicity ($H$) as
\begin{align}
    E&=\int \left[\frac{1}{2}(v_\perp^2+b_\perp^2)\right] d\tau, \,\text{and}\\
    H&=\int \vp\cdot\bp\hspace{2pt} d\tau,
\end{align}
respectively. Using the Eqs. (\ref{zmc})-(\ref{zm2}), in the absence of large-scale forcing and small-scale dissipation, one can show
{\small
\begin{align}
    \frac{d E}{d t}
    &= \int \bnabla_\perp\cdot[-\left(\frac{v_\perp^2}{2}+\frac{b_\perp^2}{2}+p^*\right)\vp+(\vp\cdot\bp)\bp]d\tau\nonumber\\
    &\hspace{100pt}+\bnabla\cdot b_0\left(\vp\cdot\bp\right)\hat{z} \,\,\,\,d\tau\nonumber\\
    &\hspace{-20pt}= \int \bnabla\cdot[-\left(\frac{v_\perp^2}{2}+\frac{b_\perp^2}{2}+p^*\right)\vp
    +(\vp\cdot\bp)\bb]\,\, d\tau,\\
   \frac{d H}{d t}
    &=\int \{\bnabla_\perp\cdot[-(\vp\cdot\bp)\vp-p^*\bp+\left(\frac{v_\perp^2}{2}+\frac{b_\perp^2}{2}\right)\bp]\nonumber\\
    &\hspace{100pt}+\bnabla\cdot b_0 \left(\frac{v_\perp^2}{2}+\frac{b_\perp^2}{2}\right)\hat{z}\}\,\,d\tau\nonumber\\
    &\hspace{-20pt}=\int \bnabla\cdot[-(\vp\cdot\bp)\vp-p^*\bp
    +\left(\frac{v_\perp^2}{2}+\frac{b_\perp^2}{2}\right)\bb]\,\,d\tau,
\end{align}}

\noindent where, for a vector $\bm{M}_\perp$ confined in the perpendicular plane, $\bnabla\cdot\bm{M}_\perp=\bnabla_\perp\cdot\bm{M_\perp}$. 
Finally, using the Gauss-divergence theorem with vanishing or periodic boundary conditions, one can show that the total energy and cross helicity are the inviscid invariants of RMHD. Therefore, both for energy and cross helicity, we can expect an inertial range cascade with constant flux.  

\section{Derivation of the exact relations}\label{exact relation}
\subsection{Exact relations in divergence form}\label{ex}
Here, we derive the exact relation corresponding to the inertial range energy transfer of the statistically homogeneous RMHD. The symmetric two-point correlator can be defined as \citep{podesta2008, banerjee2016alternative} 
\begin{small}
    \begin{align}
    \R_E&=\R_E'=\frac{1}{2}\la\vp\cdot\vp'+\bp\cdot\bp'\ra,\\
    \R_H&=\R_H'=\frac{1}{2}\la\vp\cdot\bp'+\vp'\cdot\bp\ra, 
\end{align}
\end{small}

\noindent where the unprimed and primed quantities represent the corresponding fields at $\bm{x}$ and $\bm{x+r}$ respectively and the angular bracket $\la(\cdot)\ra$ represents the ensemble average. We now calculate the evolution equation for the energy correlator ($\R_E$). Similar to the Eqs. (\ref{zmc}) and (\ref{zmi}), one can write the evolution equations for the $\vp'$ and $\bp'$ and finally obtain
{\small
\begin{align}
    &\partial_t \la\vp\cdot\vp'\ra\nonumber\\
    &=\la\vp\cdot\left[-(\vp'\cdot\bnabla_\perp')\vp'-\bnabla_\perp' {p^{*}}'+(\bp'\cdot\bnabla_\perp')\bp'\right]\ra\nonumber\\
    &+\la\vp'\cdot\left[-(\vp\cdot\bnabla_\perp)\vp-\bnabla_\perp p^{*}+(\bp\cdot\bnabla_\perp)\bp\right] \ra\nonumber\\
    &+\la\vp\cdot\left(b_0\frac{\partial \bp'}{\partial z'}\right)+\vp'\cdot\left(b_0\frac{\partial \bp}{\partial z}\right)\ra+D_u+F_u\nonumber\\
    &= \la-\bnabla_\perp'\cdot(\vp\cdot\vp')\vp'+\bnabla'\cdot(\bp'\cdot\vp) \bb'\ra\nonumber\\
    &\hspace{-1pt}+\la-\bnabla_\perp\cdot(\vp\cdot\vp')\vp+\bnabla\cdot(\bp\cdot\vp') \bb\ra+D_u+F_u\nonumber\\
    &=\bnabla_{ \bm{r}_\perp}\cdot\la-(\vp\cdot\vp')\vp'+(\vp\cdot\vp')\vp\ra\nonumber\\
    &+\bnabla_{\bm{r
    }}\cdot\la(\bp'\cdot\vp)\bb'-(\bp\cdot\vp')\bb\ra+D_u+F_u,\label{ec}
\end{align}}

\noindent where $D_u=\la\nu\left[(\vp\cdot\bnabla_{\perp}^{'2}\vp')+(\vp'\cdot\bnabla_{\perp}^2\vp)\right]\ra$ and $F_u=\la\vp\cdot\bm{f}'+\vp'\cdot\bm{f}\ra$ indicate the average dissipation and forcing contributions respectively. Finally, in the above derivation, assuming statistical homogeneity, we used $\bnabla\cdot\la(\cdot)\ra=-\bnabla_{\bm{r}}\cdot\la(\cdot)\ra=-\bnabla'\cdot\la(\cdot)\ra$ and $\la\vp\cdot\bnabla'_\perp p'^*\ra = \la\vp'\cdot\bnabla_\perp p^*\ra = 0$. Similar to Eq. (\ref{ec}), one can also obtain the evolution of $\la\bp\cdot\bp'\ra$ as 
{\small
\begin{align}
    &\partial_t\la\bp\cdot\bp'\ra\nonumber\\
    &=\la\bp\cdot\left[-(\vp'\cdot\bnabla_\perp')\bp'+(\bp'\cdot\bnabla_\perp')\vp'+b_0\frac{\partial \vp'}{\partial z'}\right]\ra\nonumber\\
    &+\la\bp'\cdot\left[-(\vp\cdot\bnabla_\perp)\bp+(\bp\cdot\bnabla_\perp)\vp+b_0\frac{\partial \vp}{\partial
     z}\right]\ra+D_m\nonumber\\
     &=\la-\bnabla_\perp'\cdot(\bp'\cdot\bp)\vp'-\bnabla_\perp\cdot(\bp\cdot\bp')\vp\ra\nonumber\\
     &+\la\bnabla'\cdot(\vp'\cdot\bp)\bb'+\bnabla\cdot(\vp\cdot\bp')\bb\ra+D_m\nonumber\\
     &=\bnabla_{ \bm{r}_\perp}\cdot\la-(\bp\cdot\bp')\vp'+(\bp\cdot\bp')\vp\ra\nonumber\\
     &+\bnabla_{\bm{r}}\cdot\la(\vp'\cdot\bp)\bb'-(\vp\cdot\bp')\bb\ra+D_m,\label{ei}
\end{align}}

\noindent
where $D_m=\la\eta(\bp\cdot\bnabla_{\perp}^{'2}\bp'+\bp'\cdot\bnabla_{\perp}^2\bp)\ra$. Adding the Eqs. (\ref{ec}) and (\ref{ei}), one can obtain the evolution of the energy correlator ($\R_E$) as
{\small
\begin{align}
    &\partial_t \R_E\nonumber\\&=\frac{1}{2}\bnabla_{\bm{r}_\perp}\cdot\la(-\vp\cdot\vp'-\bp\cdot\bp')\delta\vp\ra\nonumber\\
    &+\frac{1}{2}\bnabla_{\bm{r}}\cdot\la(\bp'\cdot\vp+\bp\cdot\vp')\delta\bb\ra+D_E+F_E,\label{re1}
\end{align}}

\noindent
where $D_E=(D_u+D_m)/2$, $F_E=F_u/2$, $\delta\vp=\vp'-\vp$ 
and $\delta\bb=\bb'-\bb$. Again, $\delta\bb=\bb'-\bb=b_0\hat{z}+\bp'-b_0\hat{z}-\bp=\delta \bp$ and hence, $\bnabla_{\bm{r}}\cdot\la(\cdot)\delta\bb\ra= \bnabla_{ \bm{r}_\perp}\cdot\la(\cdot)\delta\bp\ra$. We can therefore finally rewrite the Eq. (\ref{re1}) as
{\small
\begin{align}
    \partial_t \R_E&=\frac{1}{2}\bnabla_{ \bm{r}_\perp}\cdot\left \langle(-\vp\cdot\vp'-\bp\cdot\bp')\delta\vp \right. \nonumber\\
    &\left. +\,(\bp'\cdot\vp+\bp\cdot\vp')\delta\bp\right\rangle+D_E+F_E\nonumber\\
    &=\frac{1}{4}\bnabla_{ \bm{r}_\perp}\cdot\left\langle(\delta \vp^2+\delta \bp^2)\delta\vp-2(\delta\vp\cdot\delta\bp)\delta\bp\right\rangle\nonumber\\
    &+D_E+F_E
    \label{re3},
\end{align}}

\noindent where to obtain the last step, we used the fact that the terms $\bnabla_{ \bm{r}_\perp}\cdot\left \langle(\vp\cdot\vp)\vp'\right\rangle$, $\bnabla_{ \bm{r}_\perp}\cdot\left \langle(\bp\cdot\bp)\vp'\right\rangle$, $\bnabla_{ \bm{r}_\perp}\cdot\left \langle(\vp\cdot\bp)\bp'\right\rangle$ vanish under the assumptions of statistical homogeneity and incompressibility. Now, we consider a statistical stationary state where the left-hand side of the Eq. (\ref{re3}) vanishes. In the inertial range, we can neglect $D_E$ to finally obtain
{\small
\begin{align}
    \frac{1}{4}\bnabla_{\bm{r}_\perp}\cdot\la(\delta \vp^2+\delta \bp^2)\delta \vp-2(\delta\vp\cdot\delta\bp)\delta \bp\ra=-\varepsilon\label{re4},
\end{align}}

\noindent where the stationary state energy injection rate ($F_E$) is assumed to be equal to the energy cascade rate ($\varepsilon$) in the inertial range. In the case of cylindrical symmetry, integrating the above equation in a circle of radius $r_\perp$, we obtain
{\small
\begin{align}
    \la(\delta \vp^2+\delta \bp^2)\delta v_{r_\perp}-2(\delta\vp\cdot\delta\bp)\delta b_{r_\perp}\ra=-2\varepsilon r_\perp.\label{re5}
\end{align}}

\noindent
At this step, the final exact relation looks very similar to the exact relation derived by Politano-Pouquet (hereafter PP98) from three-dimensional incompressible MHD equations \citep{politano1998karman}. For many readers, in fact, it can be tempting to derive the above exact relation (\ref{re4}) simply by neglecting the parallel fluctuations in PP98. This has indeed been done by \citet{boldyrev2009dynamic} neglecting the parallel fluctuations in the present of a strong $\bm{b_0}$. However, following the original derivation of RMHD equations from the incompressible MHD, it turns up that in the incompressible MHD framework, the parallel fluctuations may be of the same order as that of the perpendicular fluctuations and thus can not be omitted in a straightforward way \citep{oughton2017reduced2dor3d}. Interestingly, the strongly anisotropic form of PP98, in principle, does contain both the parallel and perpendicular fluctuations and can be written as 
{\small
\begin{align}
    &\bnabla_{\bm{r}_\perp}\cdot\left\langle(\delta \vp^2+\delta\bp^2)\delta\vp-2(\delta\vp\cdot\delta\bp)\delta\bp \right.\nonumber\\
    & \left.+(\delta v_\parallel^2+\delta b_\parallel^2)\delta\vp-2(\delta v_\parallel \delta b_\parallel)\delta\bp\right\rangle=-4\varepsilon,\label{pp1}
\end{align}}

\noindent
where $\delta v_\parallel$ and $\delta b_\parallel$ are the velocity and magnetic field fluctuations along $\bm{b_0}$, respectively. However, the standard RMHD equations are derived from nearly incompressible MHD equations for which one can get rid of the parallel fluctuations. This entire discussion justifies the need of deriving the exact relation of RMHD directly from the governing equations and not as a straightforward limit of PP98. 

Similar to energy, one can also derive an exact relation for the inertial range cascade of cross helicity. Following the same methodology as that of the energy correlator, here one obtains 
{\small
\begin{align}
    &\partial_t \la \vp\cdot\bp'\ra\nonumber\\
    &=\bnabla_{ \rb_\perp}\cdot\la-(\vp\cdot\bp')\vp'+(\bp'\cdot\vp)\vp\ra\label{ch1}\nonumber\\
    &+\bnabla_{\rb} \cdot \la
    (\vp\cdot\vp')\bb'-(\bp\cdot\bp')\bb\ra+ D_1 + F_1 ,\\
    & \partial_t \la\vp'\cdot\bp\ra\nonumber\\
    &=\bnabla_{ \rb_\perp}\cdot\la(\vp'\cdot\bp)\vp-(\bp\cdot\vp')\vp'\ra\label{ch2}\nonumber\\
    &+\bnabla_{\rb} \cdot \la-(\vp\cdot\vp')\bb+(\bp\cdot\bp')\bb'\ra+ D_2+F_2,
\end{align}}

\noindent where $D_1=\la \eta\vp\cdot\bnabla_\perp^{'2} \bp' +\nu \bp'\cdot\bnabla_\perp^2\vp\ra$, $F_1=\la
\bp'\cdot\bm{f}\ra$, $D_2=\la\eta\vp'\cdot\bnabla_\perp^{2} \bp +\nu \bp\cdot\bnabla_\perp^{'2}\vp'\ra$, and $F_2=\la
\bp\cdot\bm{f}'\ra$. 
Adding the above equations, we obtain
{\small
\begin{align}
    \partial_t\R_H
    =&-\frac{1}{4}\bnabla_{ \bm{r}_\perp}\cdot\la(\delta \vp^2+\delta \bp^2)\delta \bp-2(\delta\vp\cdot\delta\bp)\delta \vp\ra\nonumber\\
    &+D_H+F_H,\label{ch3}
\end{align}}

\noindent where $D_H=(D_1+D_2)/2$ and $F_H=(F_1+F_2)/2$ and the terms $\bnabla_{ \bm{r}_\perp}\cdot\left \langle(\vp\cdot\vp)\bp'\right\rangle$, $\bnabla_{ \bm{r}_\perp}\cdot\left \langle(\bp\cdot\bp)\bp'\right\rangle$, $\bnabla_{ \bm{r}_\perp}\cdot\left \langle(\vp\cdot\bp)\vp'\right\rangle$ vanish under the assumptions of statistical homogeneity and incompressibility. Following the usual assumptions of statistical stationarity, and neglecting the dissipation term ($D_H$) in the inertial range, we obtain
{\small
\begin{align}
    \frac{1}{4}\bnabla_{ \bm{r}_\perp}\cdot\la(\delta \vp^2+\delta \bp^2)\delta \bp-2(\delta\vp\cdot\delta\bp)\delta \vp\ra=\varepsilon_H,\label{ch6}
\end{align}}

\noindent where the stationary state helicity injection rate ($F_H$) is assumed to be equal to the helicity cascade rate ($\varepsilon_H$) in the inertial range. Assuming axisymmetry with respect to $\bm{b_0}$, we finally obtain  
{\small
\begin{align}
    \la ( \delta\vp^2+\delta \bp^2)\delta b_{r_\perp}-2(\delta\vp\cdot\delta\bp)\delta v_{r_\perp}\ra=2\varepsilon_H r_\perp.\label{ch4}
\end{align}}

\noindent By adding and subtracting the Eqs. (\ref{re4}) and (\ref{ch6}), and rewriting the resultant expressions in terms of the Els{\"a}sser variables $\bm{z_\perp}^\pm=\vp\pm\bp$, we get  
{\small
\begin{align}
    \bnabla_{\bm{r_\perp}}\cdot \la(\delta\bm{z}_\perp^\pm)^2\delta \bm{z}^\mp\ra=-4\varepsilon^\pm,\label{ch5}
\end{align}}

\noindent where $\varepsilon^\pm=\varepsilon\pm\varepsilon_H$ denote the mean cascade rates of the pseudo energies ($E^\pm=\frac{1}{2}\bm{z}^\pm \cdot \bm{z}^\pm$), which are equivalently conserved in inviscid incompressible MHD. Similarly, assuming cylindrical symmetry, one obtains
{\small
\begin{align}
    \la(\delta\bm{z}_\perp^\pm)^2\delta z^\mp_{r_\perp}\ra=-2\varepsilon^\pm r_\perp.\label{ch7}
\end{align}}

\subsection{Energy spectra in RMHD turbulence}

Using the exact relations derived above, one can predict the anisotropic energy spectra in RMHD turbulence. In the case where strong alignment is not considered, one can have $\delta z^+_\perp \sim \delta z^-_\perp\sim \delta v_\perp \sim \delta b_\perp$, and using Eq. (\ref{re5}), also obtain
\begin{equation}
    v_{r_\perp}^3 \sim \varepsilon r_\perp,
\end{equation}
where $v_{r_\perp}=\sqrt{(\delta v_\perp)^2}$. Defining anisotropic energy spectra $E(k_\perp, k_\parallel)k_\perp k_\parallel\sim v_{r_\perp}^2 $ and assuming $\varepsilon$ to be scale-independent, we have 
{\small
\begin{align}
     &E(k_\perp, k_\parallel)k_\perp k_\parallel \sim \varepsilon^{2/3} k_\perp^{-2/3}\\
    \Rightarrow \ &E(k_\perp, k_\parallel) \sim k_\perp^{-5/3}k_\parallel^{-1},\label{ch8}
\end{align}}

\noindent which is the anisotropic energy spectra consistent with critical balance and naturally satisfies the relation $3\alpha+2\beta=7$ with $\alpha = 5/3$ and $\beta = 1$, where $E(k_\perp, k_\parallel)\sim k_\perp^{-\alpha} k_\parallel^{-\beta}$\citep{galtier2005spectral}.

This result, however, does not hold if there is a strong alignment between $\delta \vp$ and $\delta \bp$. In addition to the parallel fluctuation scale $l$, there we have two distinct length scales $\lambda$ and $\xi$ in the plane perpendicular to $\bm{b_0}$ where $\lambda$ is along the direction of alignment and $\lambda\ll\xi\ll l$ is satisfied. In particular, the scale-dependent alignment between the fluctuations $\bm{v}_{ \lambda}$ and $ \bm{b}_\lambda$ is given by the small angle $\theta_\lambda \sim \lambda/\xi \sim \lambda^{1/4}$ \citep{Boldyrev_2005, boldyrev2006}. By straightforward reasoning, one can therefore relate the Els{\"a}sser variables as \citep{boldyrev2009dynamic}
\begin{align}
    z^+_\lambda\sim  v_\lambda,\,\,\, 
    z^-_\lambda\sim v_\lambda \theta_\lambda.
\end{align}
Using the above relation, from Eq. (\ref{ch5}), we dimensionally have 
{\small
\begin{equation}
     v_\lambda^3 \theta_\lambda\sim \varepsilon^+\lambda
     \Rightarrow v_\lambda\sim (\varepsilon^{+})^{1/3}\lambda^{1/4}.
\end{equation}}

\noindent Defining three-dimensionally anisotropic energy spectra $E(k_\lambda, k_\xi, k_\parallel)k_\lambda k_\xi k_\parallel\sim  v_\lambda ^2$ and assuming $\varepsilon^+$ to be scale-independent, one obtains
{\small
\begin{align}
    &E(k_\lambda, k_\xi, k_\parallel)k_\lambda k_\xi k_\parallel\sim k_\lambda^{-1/2}\\
    \Rightarrow& E(k_\lambda, k_\xi, k_\parallel)\sim k_\lambda^{-3/2} k_\xi^{-1} k_{\parallel}^{-1},\label{ch9}
\end{align}}

\noindent which consists of three exponents and therefore, is different from Eq. (\ref{ch8}). Note that, the above energy spectra is consistent with the reduced energy spectra $E(k_\lambda)\sim k_\lambda^{-3/2}$, $E(k_\xi)\sim k_\xi^{-5/3}$ and $E( k_\parallel)\sim k_\parallel^{-2}$ obtained by \citet{boldyrev2006}. However, it is possible to find a consecutive relation for the three exponents obtained above. The derivation fits as a separate study to be submitted elsewhere (Sasmal and Banerjee, to be submitted).

\section{Alternative exact relations and turbulent relaxed states}\label{alternative exact relation}

In order to obtain the alternative form of the exact relations, let us obtain, for an arbitrary vector $\bm{s}$, an expression for $(\bm{s_\perp}\cdot\bnabla_\perp)\bm{s}_\perp$. For a vector $\bm{s_\perp}$, we have
{\small
\begin{align}
    &\bnabla \left(\frac{s_\perp^2}{2}\right)= (\bm{s_\perp}\cdot\bnabla)\bm{s_\perp}+\bm{s_\perp}\times(\bnabla\times \bm{s_\perp})\nonumber\\
    &=(\bm{s_\perp}\cdot\bnabla_\perp)\bm{s_\perp}+\bm{s_\perp}\times(\bnabla\times \bm{s_\perp})\nonumber\ (\because \bm{s_\perp}\cdot\hat{z}\frac{\partial}{\partial z}=0)\nonumber\\
    &\Rightarrow(\bm{s_\perp}\cdot\bnabla_\perp)\bm{s_\perp} =\bnabla \left(\frac{s_\perp^2}{2}\right)-\bm{s_\perp}\times(\bnabla\times \bm{s_\perp})\nonumber\\
    &=\bnabla_\perp \left(\frac{s_\perp^2}{2}\right)+\frac{\partial }{\partial z}\left(\frac{s_\perp^2}{2}\right)\hat{z}-\bm{s_\perp}\times[(\bnabla_\perp+\frac{\partial}{\partial z}\hat{z})\times\bm{s_\perp}]\nonumber\\
    &=\bnabla_\perp \left(\frac{s_\perp^2}{2}\right)-\bm{s_\perp}\times(\bnabla_\perp\times \bm{s_\perp})\\
    &\because \frac{\partial }{\partial z}\left(\frac{s_\perp^2}{2}\right)\hat{z}-\bm{s_\perp}\times(\frac{\partial}{\partial z}\hat{z}\times \bm{s_\perp})=0\nonumber.
\end{align}}
\noindent Using the above identity for $\vp$ and $\bp$, one can obtain 
{\small
\begin{align}
    (\vp\cdot\bnabla_\perp)\vp&=\bnabla_{\perp} (\vp^2/2)- \vp\cross(\bnabla_\perp\cross\vp),\\
    (\bp\cdot\bnabla_\perp)\bp&=\bnabla_{\perp} (\bp^2/2)- \bp\cross(\bnabla_\perp\cross\bp).
\end{align}}

\noindent Finally, substituting the above expressions in Eqs. (\ref{zmc}) and (\ref{zmi}), we write the equations of RMHD as 
{\small
\begin{align}
     \frac{\partial \vp}{\partial t}&=\vp\cross\wz-\bnabla_\perp P-\bp\cross\jz+b_0\frac{\partial \bp}{\partial z}+\bm{d}+\bm{f},\label{af1}\\
    \frac{\partial \bp}{\partial t}&=\bnabla_\perp\cross(\vp\cross\bp)+b_0\frac{\partial \vp}{\partial z}+\bm{d_b},\label{af2}
\end{align}}

\noindent where $\wz=\bnabla_\perp\cross\vp$, $P=p+\frac{\vp^2}{2}$, $\jz=\bnabla_\perp\cross\bp$, $\bm{d}$ and $\bm{d_b}$ represent dissipation and $\bm{f}$ represents forcing. Following \citet{banerjee2016alternative}, one can obtain the evolution of $\la\vp\cdot\vp'\ra$ as 
{\small
\begin{align}
    &\partial_t \la\vp\cdot\vp'\ra\nonumber\\
    &=\la\vp\cdot\left(\vp'\cross\wz'-\bnabla_\perp' P'-\bp'\cross\jz'+b_0\frac{\partial \bp'}{\partial z'}\right)\ra\nonumber
    \\&+\la\vp'\cdot\left(\vp\cross\wz-\bnabla_\perp P-\bp\cross\jz+b_0\frac{\partial \bp}{\partial z}\right) \ra+D_u+ F_u\nonumber\\
    &=\la \vp \cdot\left(\vp'\cross\wz'-\bp'\cross\jz'\right)+ \vp'\cdot\left(\vp\cross\wz-\bp\cross\jz\right)\ra\nonumber\\
    &+b_0\la \frac{\partial }{\partial z'}(\bp'\cdot\vp)+\frac{\partial }{\partial z}(\bp\cdot\vp')\ra+D_u+ F_u\nonumber\\
    &=\la\delta \vp\cdot\delta\left[-(\vp\cross\wz)+(\bp\cross\jz)\right]-2\vp\cdot(\bp\cross\jz)\ra\nonumber\\
    &+b_0\bnabla_{\bm{r}}\cdot\la (\bp'\cdot\vp-\bp\cdot\vp')\hat{z}\ra+D_u+ F_u,\label{af3}
\end{align}}

\noindent where we have used $\vp\cdot(\vp\times\wz)=0$, and the terms involving pressure gradient vanish due to homogeneity and incompressibility.
Similarly, one can obtain
{\small
\begin{align}
    &\partial_t\la\bp\cdot\bp'\ra\nonumber\\
    &=\la\bp\cdot\lbracket\bnabla'_\perp\cross(\vp'\cross\bp')+b_0\frac{\partial \vp'}{\partial z'}\rbracket\ra\nonumber\\
    &+\la\bp'\cdot\lbracket\bnabla_\perp\cross(\vp\cross\bp)+b_0\frac{\partial \vp}{\partial z}\rbracket\ra+D_m\nonumber\\
    &=\la\jz\cdot(\vp'\cross\bp')+\jz'\cdot(\vp\cross\bp)\ra\nonumber\\
    &+b_0\la\frac{\partial }{\partial z'}(\vp'\cdot\bp)+\frac{\partial }{\partial z}(\vp\cdot\bp')\ra+D_m\nonumber\\
    &=-\la\delta \jz \cdot\delta(\vp\cross\bp)-2
    \jz\cdot(\vp\cross\bp)\ra\nonumber\\
    &-b_0\bnabla_{\bm{r}}\cdot\la(\bp'\cdot\vp-\bp\cdot\vp')\hat{z}\ra\label{af4}+D_m.
\end{align}}

\noindent Adding the Eqs. (\ref{af3}) and (\ref{af4}), one obtains the evolution of energy correlator as
{\small
\begin{align}
    &\partial_t \R_E\nonumber\\
    &=\frac{1}{2}\la\delta \vp\cdot\delta(\wz\cross\vp+\bp\cross\jz)
    -\delta \jz \cdot\delta(\vp\cross\bp)\ra\nonumber\\
    &+D_E+F_E.\label{af5}
\end{align}}

\noindent Finally, to obtain the exact relation, we consider a stationary state where $\partial_t \R_E=0$. In addition, we also consider the length scales to be well inside the inertial range where the dissipative term can be neglected and the energy injection rate is determined by the forcing term i.e. $F_E=\varepsilon$. The resulting exact relation can be written as
{\small
\begin{align}
    \la\delta \vp\cdot\delta(\vp\cross\wz+\jz\cross\bp)
    +\delta \jz \cdot\delta(\vp\cross\bp)\ra=2\varepsilon,\label{af6}
\end{align}}

\noindent which is the alternative form for the exact relation of inertial range energy transfer in RMHD turbulence. To derive the alternative form of the exact relation for the cross helicity transfer, we calculate the evolution of $\la\vp\cdot\bp'\ra$ and $\la\vp'\cdot\bp\ra$ as 
{\small
\begin{align}
    &\partial_t\la\vp\cdot\bp'\ra\nonumber\\
    &=\la\wz\cdot(\vp'\cross\bp')+ \bp' \cdot\left(\vp\cross\wz
    -\bp\cross\jz\right)\ra \nonumber\\
    &+b_0\bnabla_{\bm{r}}\cdot\la(\vp'\cdot\vp-\bp\cdot\bp')\hat{z}\ra+D_1+F_1,\\
    &\partial_t\la\vp'\cdot\bp\ra\nonumber\\
    &=\la\wz'\cdot(\vp\cross\bp)+ \bp \cdot\left(\vp'\cross\wz'
    -\bp'\cross\jz'\right)\ra\nonumber\\
    &-b_0\bnabla_{\bm{r}}\cdot\la(\vp'\cdot\vp-\bp\cdot\bp')\hat{z}\ra+D_2+F_2,
\end{align}}

\noindent respectively. Adding the above two equations and using the similar properties stated above, one obtains the alternative form of the evolution of cross-helicity correlator as
{\small
\begin{align}
    &\partial_t \R_{H}\nonumber\\
    &=\frac{1}{2}\la\delta\bp\cdot\delta(\wz\cross\vp
    +\bp\cross\jz)-\delta\wz\cdot\delta(\vp\cross\bp)\ra\nonumber\\
    &+D_H+F_H\label{alth}
\end{align}}
In the statistical stationary state, using similar assumptions as stated in \ref{ex}, we get the alternative form of the Eq. (\ref{ch6}) as 
{\small
\begin{align}
    \hspace{-2pt}\la\delta\wz\cdot\delta(\vp\cross\bp)+\delta\bp\cdot\delta
    (\vp\cross\wz
    +\jz\cross\bp)\ra
    =2\varepsilon_H.\label{af7}
\end{align}}
Although, the Eqs. (\ref{af6}) and (\ref{af7}) look similar to the exact relations derived in \citet{banerjee2016alternative}, the above equations can not be recovered as the straightforward anisotropic version of the latter due to the same reason discussed after Eq. (\ref{pp1}). 

\subsection{Relaxed states of RMHD}
In this section, we shall investigate the states to which a fully developed turbulent RMHD flow relaxes when the turbulence forcing is quenched. For obtaining the turbulent relaxed states, we use the recently developed principle of vanishing nonlinear transfer (PVNLT), according to which, the average scale to scale nonlinear transfers of inviscid invariants vanish \citep{banerjee2023universal}. The average nonlinear transfers of the total energy and cross-helicity are given by
{\small
\begin{align}
    &\la \F^E_{tr} \ra\nonumber\\
    &=  \frac{1}{2}\left \langle \vp \cdot\left(\vp'\cross\wz'-\bnabla_\perp' P'-\bp'\cross\jz'\right)\right.  \nonumber\\
    &+ \vp'\cdot\left(\vp\cross\wz-\bnabla_\perp P-\bp\cross\jz\right)\nonumber\\
    &+\left.\jz\cdot(\vp'\cross\bp')+\jz'\cdot(\vp\cross\bp)\right\rangle,\\
    &\la \F^{H}_{tr} \ra\nonumber\\
    &=\frac{1}{2}\left\langle\bp \cdot\left(\vp'\cross\wz'-\bnabla_\perp' P'-\bp'\cross\jz'\right)\right.\nonumber\\
    &+ \bp'\cdot\left(\vp\cross\wz-\bnabla_\perp P-\bp\cross\jz\right)\nonumber\\
    &+\left.\wz\cdot(\vp'\cross\bp')+\wz'\cdot(\vp\cross\bp)\right\rangle,
\end{align}}

\noindent respectively. For a turbulent relaxed state, PVNLT requires $\la \F^E_{tr}\ra=0$ and $\la \F^H_{tr}\ra=0$ which can be achieved if
\begin{align}
    \vp\cross\wz-\bp\cross\jz&=\bnabla_\perp (P+\phi_0),\label{rs1}\\
    \vp\cross\bp&=\bnabla_\perp \psi_0,\label{rs2}
\end{align}
where $\phi_0$ and $\psi_0$ are arbitrary  system-specific scalar fields. 
By careful inspection, one can note that in Eq. (\ref{rs2}), the vector $\vp\times\bp$ is along the $z$ direction whereas $\bnabla_\perp \psi_0$ lies in the $x-y$ plane. Hence, Eq. (\ref{rs2}) practically becomes
\begin{align}
    \vp\times\bp=0\label{rs4}.
\end{align}
\noindent From Eq. (\ref{rs4}), we have $\vp=\lambda\bp$ which,   assuming $\lambda$ to be constant, also gives $ \wz=\lambda\jz$ . Finally, assuming $\bnabla_\perp \phi_0$ to vanish, we have
\begin{align}
    \vp\cross\wz-\bp\cross\jz&=\bnabla_\perp P,\label{rs3}
\end{align}
and putting the above results in this equation, we get
\begin{equation}
    (1-\lambda^2)(\jz\cross\bp)=\bnabla_\perp (p+\frac{\vp^2}{2}).
\end{equation}

\noindent Under the situation where the pressure gradient is negligibly small, the above equation becomes
\begin{align}
    (\jz\cross\bp)=\frac{\lambda^2}{1-\lambda^2}\bnabla_\perp \frac{\bp^2}{2}.\label{rs5}
\end{align}

\noindent By construction, in RMHD turbulence, $\vert\vp\vert\sim \vert\bp\vert$ and hence, $\lambda$ can not be very small. This is consistent with the fact that $\bm{j}\times\bp\neq\bm{0}$ which is trivially satisfied for RMHD. Now, in the case where $\bnabla_\perp \bp^2/2$ is negligibly small, the system relaxes towards a state of dynamic alignment ($\lambda=\pm1$) which is similar to a 3D incompressible MHD flow starting with a very low magnetic helicity \citep{stribling1991relaxation}. 

\section{Comparison with two-dimensional magnetohydrodynamics}\label{comparision}
Albeit certain similarities, RMHD is subtly different from incompressible two-dimensional magnetohydrodynamics. The governing equations of incompressible 2D magnetohydrodynamics are written as
 {\small
 \begin{align}
     \hspace{-8pt}\frac{\partial \vp}{\partial t} + \vp\cdot\bnabla_\perp \vp &= -\bnabla_\perp p^*+\bp\cdot\bnabla_\perp \bp+ \bm{f} +\bm{d}, \label{mh1}\\
    \frac{\partial \bp}{\partial t}+\vp\cdot\bnabla_\perp \bp&=\bp\cdot\bnabla_\perp \vp+\bm{d_b}, \label{mh2}\\
    \bnabla_\perp\cdot\vp&=0,\;\;\bnabla_\perp\cdot\bp=0,\label{mh3}
 \end{align}}
 
\noindent where, ${\bf \vp}$ and ${\bf \bp}$ are the entire velocity and magnetic field vectors (as the flow is constrained solely in $x-y$ plane), $\bm{f}$, $\bm{d}$, and $\bm{d_b}$ represent the usual forcing and dissipation terms, respectively. Unlike RMHD, there is no z-dependence in 2D MHD, and hence the terms $\left({\partial\vp}/{\partial z} \right)$ and $ \left({\partial\bp}/{\partial z} \right)$ vanish identically in 2D MHD. Furthermore, in addition to energy ($E$) and cross helicity ($H$), it also conserves mean square magnetic potential $ (A =\int a^2 \hspace{2pt} d\tau)$ which is not an inviscid invariant of RMHD \citep{Pouquet_1978_2dmhd}. The magnetic and kinetic helicities
are trivially zero as the magnetic vector potential ($\bm{a}$) and vorticity ($\bm{\omega}$) are along $\hat{z}$ direction $i.e.$ perpendicular to the plane of flow. 

We shall now compare the exact relations of 2D MHD with those obtained for RMHD. The two-point correlators for inviscid invariants of 2D MHD are defined as
 {\small
\begin{align}
    \R_E&=\R_E'=\frac{1}{2}\la\vp\cdot\vp'+\bp\cdot\bp'\ra,\\
    \R_{H}&=\R'_{H}=\frac{1}{2}\la\vp\cdot\bp'+\vp'\cdot\bp\ra,\\
    \R_{A}&=\R'_{A}=\la \bm{a}\cdot \bm{a'}\ra,
\end{align}}

\noindent where the unprimed and primed quantities have their usual meaning of representing the corresponding fields at $\bm{x}$ and $\bm{x + r}$ respectively. Following the similar mathematical steps in the section (\ref{ex}), one obtains the exact relations for the cascade of the invariants in the inertial range of statistically homogeneous 2D MHD as
{\small
\begin{align}
    &\hspace{-6pt}\bnabla_{ \bm{r}_\perp}\cdot\la\frac{\delta \vp}{4}(\delta \vp^2+\delta \bp^2)-\frac{\delta \bp}{2}(\delta\vp\cdot\delta\bp)\ra=-\varepsilon,\label{2dmhde1}\\
    &\hspace{-6pt}\bnabla_{\bm{r}_\perp}\cdot\la\frac{\delta \bp}{4}(\delta \vp^2+\delta \bp^2)-
    \frac{\delta \vp}{2}(\delta\vp\cdot\delta\bp)\ra=\varepsilon_H,\label{2dmhde2}\\
    &\hspace{-6pt}\bnabla_{\rb_\perp}\cdot\la\frac{\delta\vp}{2}(\delta \bm{a})^2\ra =-\varepsilon_A,\label{2dmhde3}
\end{align}}

\noindent where $\varepsilon$, $\varepsilon_H$, and $\varepsilon_A$ represent the flux rates of energy, cross-helicity, and mean square magnetic potential respectively in the inertial range. Whereas the exact relations (\ref{2dmhde1}) and (\ref{2dmhde2}) are exactly similar to those obtained in RMHD, the exact relation in Eq. (\ref{2dmhde3}) has no counterpart in RMHD turbulence. For energy and cross-helicity cascade in 2D MHD, one can also obtain alternative forms of exact relations which are identical to those in RMHD. For the third invariant $A$, the alternative form of the exact relation is given by $\la\delta\bm{a}\cdot\delta(\vp\times\bp)\ra=\varepsilon_A$. Using PVNLT, one can also show that the turbulent relaxed states in 2D MHD are identical to Eqs. (\ref{rs4}) and (\ref{rs3}). Unlike RMHD, in 2D MHD, a priori we can not put any restriction on the value of $\lambda$. However, if we assume $\lambda\ll1$ and the gradient of pressure is also negligibly small, then, the relaxation condition reduces to $(1-\lambda^2)(\jz\times\bp)=0\Rightarrow \lambda=\pm1$ (as $\jz\perp\bp$), leading to an inconsistency. The relaxed states of 2D MHD will therefore be similar to Eq. (\ref{rs5}) with $\lambda=\pm1$ as one of the possible solutions.

\section{Discussion}\label{discussion}
In this paper, starting from the basic equations of RMHD, we have derived exact relations for inertial range energy and helicity transfer in fully developed homogeneous turbulence in fusion plasmas. In order to keep consistency with the inherent near-incompressible (but not incompressible) nature of RMHD, we consciously avoided the `wormhole' derivation of the said exact relations as a strong anisotropic limit of the exact relation derived for incompressible MHD turbulence where we cannot unconditionally eliminate the parallel fluctuations of velocity and magnetic field. Using the exact relations, we have also  predicted an anisotropic energy spectra $E(k_\perp, k_\parallel) \sim k_\perp^{-5/3}k_\parallel^{-1}$, which is consistent with the prediction of critical balance and satisfies a constitutive relation $3 \alpha + 2 \beta =7$ with $\alpha$ and $\beta$ being the exponents of perpendicular and parallel wave numbers, respectively. Relaxing the axisymmetry condition in the presence of strong alignment, we have also successfully explained the emergence of a perpendicular $-3/2$ spectra in the direction of alignment. This is due to anisotropic strong cascade of energy for highly Alfv\'enic case and is probably observed in the energy spectra of the turbulent solar wind close to the sun (Mondal, Banerjee and Sorriso-Valvo, submitted to ApJ). Note that, this is constitutionally different from the Iroshnikov-Kraichnan $-3/2$ spectra which one may expect due to weak energy cascade in balanced MHD turbulence. submitted).  

In addition to the usual divergence form, we have also derived alternative exact relations for energy and cross helicity cascades. Similar to the divergence form, here also one should not be tempting to derive this as a direct anisotropic limit of the exact relation of incompressible MHD turbulence \citep{banerjee2016alternative}. Using PVNLT, we found that fully developed RMHD turbulence relaxes towards a state of dynamic alignment when the pressure gradients are negligibly small. Dynamic alignment is the natural relaxed state for a three-dimensional incompressible MHD with negligibly small magnetic helicity and hence is a reasonable relaxed state where the magnetic helicity vanishes identically at every point of the flow. However, unlike three dimensional MHD, a turbulent RMHD is shown to never relax to a state of Beltrami-Taylor alignment. Finally, despite having slightly different governing equations and one additional inviscid invariant (mean square magnetic potential), two-dimensional MHD turbulence possesses identical exact relations to those of RMHD turbulence for the energy and cross-helicity transfers. On quenching the turbulent forcing, a 2d-MHD system is also found to relax to a state of dynamic alignment similar to an RMHD flow. However, such similarity is limited to the regime of strong turbulence as an RMHD flow may experience shear Alfv\'en wave turbulence whereas a two-dimensional MHD turbulence may eventually develop a weakly sustaining non-local wave turbulence with Pseudo-Alfv\'en waves  \citep{galtier2006, tronko2013}. 

The derived exact relations can be theoretically very important and may serve as a benchmark for the large scale limit of the exact relations in gyro-kinetic fluid models. A typical example is the case of finite Larmor radius MHD turbulence which is believed to play a pivotal role in the anomalous turbulent heating of the solar wind through the mechanism of helicity barrier \citep{Meyrand2021, Squire2022}. In the realm of applications, our obtained relation can give an exact estimate for the heating by turbulent cascade in both the near sun solar wind and the fusion plasmas.  

\section{Acknowledgement} The authors acknowledge S\'ebastien Galtier and Vincent David for useful discussions. S.B. acknowledges the financial support from STC-ISRO grant (STC/PHY/2023664O).


%

\end{document}